\documentclass[english]{article}
\usepackage{filecontents}
\usepackage[T1]{fontenc}
\usepackage[latin9]{inputenc}
\usepackage{physics}
\usepackage{mathrsfs}
\usepackage{amsmath}
\usepackage{amssymb}
\usepackage{setspace}
\usepackage{authblk}
\usepackage{calrsfs}

\makeatletter

\usepackage{mathrsfs}
\usepackage{babel}

\usepackage{babel}

\makeatother

\usepackage{babel}
\begin{document}

	\title{\textbf{A note on broken dilatation symmetry in planar noncommutative theory}}

	\author{Partha Nandi\footnote{parthanandi@bose.res.in}$~^1$,~Sankarshan Sahu\footnote{sankarshan.sahu2000@gmail.com}$~^2$,~Sayan Kumar Pal \footnote{sayankpal@bose.res.in}$~^1$}
	\affil{$^1$ S.~N.~Bose National Centre for Basic Sciences,}
	\affil{JD Block, Sector III, Salt Lake, Kolkata-700106, India.}
	\affil{$^2$ Indian Institute of Engineering Science and Technology, Shibpur,}
	\affil{Howrah, West Bengal-711103, India.}
	\maketitle
	
	
	
	%
	
	\begin{abstract}

	A study of a riveting connection between noncommutativity and the anomalous dilatation (scale) symmetry is presented for a generalized quantum Hall system due to time dilatation transformations. On using the "Peierls substitution" scheme, it is shown that noncommutativity between spatial coordinates emerges naturally at a large magnetic field limit. Thereafter, we derive a path-integral action for the corresponding noncommutative quantum system and discuss the equivalence between the considered noncommutative system and the generalized Landau problem thus rendering an effective commmutative description. By exploiting the path-integral method due to Fujikawa, we derive an expression for the unintegrated scale or dilatation anomaly for the generalized Landau system, wherein the anomalies are identified with Jacobian factors arising from measure change under scale transformation and is subsequently renormalised. In fact, we derive exact expressions of anomalous Ward identities from which one may point out the existence of scale anomaly which is a purely quantum effect induced from the noncommutative structure between spatial coordinates.

	\end{abstract}
		\section{Introduction}
			Scale or dilatation symmetry has quite a long history \cite{COLEMAN1971552, Pokorski:1987ed} in particle physics and has become an interesting concept in present day physics for its usefulness in understanding a variety of physical phenomena ranging from phase transitions in statistical mechanics, stochastic processes, many-body condensed matter systems to construction of conformal quantum field theories and even in plane gravitational waves \cite{ZinnJustin:2002ru, continentino_2017, PhysRevLett.59.381, 1984NuPhB.241..333B, Zhang:2019koe}. A classical example in mechanics which respects this symmetry is the inverse square potential in $2+1$ and $3+1$ dimensions \cite{Gozzi:2005kx}. Further in $2+1$ dimensions, the delta function potential $\delta^{2}(\vec{r})$ is also a scale invariant theory \cite{Jackiw:1995be}. On the other hand in classical field theories, the massless Klein Gorden as well as massless  phi-four theory also respect this symmetry \cite{DiFrancesco:1997nk}. However, the phenomenon of conservation of certain currents or charges valid in classical domain may not continue to hold upon quantization - the appearance of anomalies. Anomaly is one of the three possible types of breaking of a symmetry exhibited by a physical system at the quantum level with the other two being classical/explicit and spontaneous symmetry breaking effects. As a matter of fact, nature does not respect scale invariance as far as particle physics is concerned \cite{COLEMAN1971552, Pokorski:1987ed}. The most famous example of an anomaly which created much debate and attention in high energy physics is the so-called ABJ anomaly and the PCAC (partial
			conservation of axial-vector current) hypothesis \cite{PhysRev.177.2426, Bell:1969ts, PhysRevLett.42.1195}. In this regard, in anomalous gauge theories, it has been shown from algebraic consistency in \cite{PhysRevLett.61.514} that the electric fields must be noncommutative.  Also it is worthwhile to mention here that the explicit structure for the chiral anomaly in non-commutative gauge theories has been already well studied in the literature \cite{PhysRevD.71.045013, Bonora:2001fa}. \\
			
			In modern physics, the study of symmetries of physical systems and also on their breaking has helped to gain deep insights into their physical content and also on their underlying mathematical (geometrical) structure. Study of anomalies has become indispensable to explore novel features in the study of quantum phase transitions, see for example \cite{Brattan:2017yzx} and references therein. Moving ahead, it has been shown by R. Jackiw in \cite{JACKIW199083} that a charged particle interacting with a vortex magnetic field enjoys the scale or time dilatation symmetry, as well as other time reparametrization symmetry namely, conformal transformation and transformation due to time translation. The invariance group of this system turns out to be $SO(2,1)\times U(1)$ and it admits a hidden $SO(2,1)$ dynamical symmetry group of conformal transformations, where generators of this dynamical symmetry are generators of the above three time reparametrization transformations. Furthermore, this leads to a deformed Heisenberg algebra i.e. noncommutative momenta as a by-product. In this context, it is important to mention here that recently one of us \cite{PhysRevA.102.022231} has shown that due to deformed planar Heisenberg algebra with phase space non-commutativity, the system Hamiltonian of a 2D harmonic oscillator can be recast as an algebra element of $SO(2,1)\times U(1)$ which is responsible for the emergence of Berry phase in such systems under adiabatic evolution of the system Hamiltonian. A relation between ABJ anomalies and occurrence of Berry phase has been unfolded in \cite{PhysRevLett.109.181602} for Fermi liquids. Again, it has been investigated in \cite{Kamath:1992bv} for the dilatation anomaly in the Landau problem. In this regard, it is interesting to notice that the problem of charged particle interacting with a point magnetic vortex differs from the Landau problem under scaling transformations even though the form of interactions is the same. Now, as a matter of fact, it is well-known now that in the Landau problem, guiding-center coordinates satisfy a noncommutative commutation relation. It is therefore quite natural 
		 to investigate the fate of scale symmetry breaking and occurrence of an anomaly, if any, in deformed (noncommutative) planar systems taking the harmonic oscillator as a first study which has been lacking in the literature so far. We will carry out the analysis both in the classical as well as in quantum domain using path-integral techniques. In this context, recently, a path-integral derivation of the nonrelativistic scale anomaly in usual commutative space has been carried out in \cite{PhysRevD.91.085023}. It may be useful to mention here that in \cite{PhysRevD.74.085032}, a connection between  axial anomaly of QED in a strong
		 	magnetic field and the non-planar axial anomaly of the  conventional non-commutative $U(1)$ gauge theory has been observed.\\

		 Deformed (noncommutative) quantum physics has gained much attention in the last two or three decades as a promising route towards constructing a consistent theory of gravity. Actually the concept of noncommutativity dates back much earlier in the works of \cite{doi:10.1098/rspa.1938.0060, PhysRev.72.874} in an attempt to unify quantum mechanics with gravity. But soon it faded out with the success of renormalization programme. Howbeit starting from 90's, different approaches in high energy physics have pointed out at the existence of a minimal length in Planck scale regime and noncommutative physics has revived from this confluence of ideas \cite{Doplicher_1995, Hooft_1996, Majid:1988we, Seiberg_1999, Aschieri_2006, PhysRevLett.111.101301}. However, the concept of noncommutativity has remained no more just a theoretical interest relevant in very high energy physics but has emerged naturally in condensed matter systems and has phenomenological consequences as well. For an example, in semi-classical dynamics of Bloch electrons exhibiting Berry curvature, the coordinates become noncommutative \cite{PhysRevLett.95.137204} and lead to anomalous Hall conductance in ferromagnetic semiconductors \cite{PhysRevLett.88.207208}. Closely related to this development, there has been other advances in noncommutative physics from the perspective of Galilean symmetry in planar nonrelativistic mechanics. In (2+1) dimensions, the Galilean algebra admits a unique second central extension which can be identified with nonrelativistic spin \cite{bose1995, Jackiw_2000} and also this can be corroborated from a consistent nonrelativistic limit of Poincare group in (2+1) dimensions. Now, it has been shown explicitly that the deformation parameter in noncommutative mechanics can be identified with the spin of particles in two dimensions (anyonic spin) \cite{Duval:2001hu, Horvathy:2010wv} 
		 and therefore noncommutative quantum mechanics helps in the understanding of spin as well as fractional quantum Hall effect \cite{Berard:2004xn, Jellal:2006dj, PhysRevLett.107.116801} where the exotic statistics of anyons play an important role. Moreover, spatial noncommutativity has been suggested to emerge in the physics of cold Rydberg atoms \cite{PhysRevLett.74.514, PhysRevLett.93.043002}. Very recently it has been reported in \cite{PhysRevLett.121.120401} that there have been anomalous frequency shifts in ultracold Rydberg atoms implying breaking of scale symmetry in such systems. This particular observation adds further to the motivation of the present study. Here in the current work, we investigate existence of anomaly under scaling transformations in deformed (noncommutative) quantum system, in particular in exotic oscillators. The form of the effective action of exotic oscillators (\ref{nwe ac}) will be reminiscent of the action of a charged particle interacting with magnetic point vortex under appropriate variables \cite{JACKIW199083}. It will be shown that the system does not admit scaling symmetry classically on account of the non-zero spring constant. Even so strikingly, at the quantum level there appears an anomalous contribution to the non-conservation of dilatation charge. This is an interesting result that we report here and to the best of our knowledge, this paper is a first direction towards the study of scale anomalies in deformed quantum systems. This paper also pursues the current problem in the spirit of a (0+1) dimensional noncommutative field theory \cite{PhysRevD.80.105014} employing methods generally used in field theories as quantum mechanics can be considered to be a (0+1) dimensional quantum field theory \cite{JACKIW1979158}.\\

		The paper has been arranged as follows. We start by demonstrating in section 2 how deformed harmonic oscillators can arise in the classical setting by considering Peierls-substitution scheme \cite{doi:10.1142/9789812795779_0010,DUNNE1993114} and show that this represents a system of electric dipole under a very strong magnetic field with a harmonic interaction between the charges along with an extra harmonic potential part of one of the charges. In section 3, we then move on to the quantum setting where we formulate the noncommutative quantum mechanics of a two-dimensional harmonic oscillator in the framework of Hilbert-Schmidt operators, an operator formulation of noncommutative quantum mechanics. Then we will construct the coherent-state path integral of the noncommutative quantum system and compute the phase-space action for the system where a resemblance with Chern-Simons quantum mechanics shall be made in section 4. Symplectic brackets can be computed with this action and the noncommutative structure is figured out between the position coordinates. In section 5, with the help of the generating functional for Green's functions, we explicitly compute Ward-Takahashi identities as a (0+1) dimensional noncommutative field theory. These identities will get plagued with anomalies as quantum corrections owing to the existence of noncommutativity. The anomaly term needs to regularized and is done in section 6 using Fujikawa's regulator. The anomaly term will be found to be proportional to the product between the spring constant of the harmmonic oscillator and the noncomutative parameter. Note that initially during the construction of generating function in presence of source, we have worked with noncommutative variables. Then we have switched to canonical phase-space variables for further development in the computation of Ward identities.
		
\section{The classical picture of noncommutative space}
		 We consider, inspired by the "Peierls-substitution" scheme, a pair of non-relativistic interacting opposite charged particles having same mass $m$ moving on the plane in a constant magnetic field $B$ along $z$ axis (within the approximation where we can ignore Coulomb and radiation effects). The coordinates of negative and positive charges are denoted in components form by $x_{i}$ and $y_{i}$( $i=1,2$) respectively. The dynamics of the system is confined in a plane and therefore the $z$ coordinate can be suppressed. The system is described by the following standard Lagrangian in C.G.S units :
		\begin{equation}
		L= \frac{1}{2}m(\dot{x}^{2}_{i}+\dot{y}^{2}_{i})+ \frac{eB}{2c}\epsilon_{ij}(x_{j}\dot{x}_{i}-y_{j}\dot{y}_{i})-\frac{k_{0}}{2}(x_{i}-y_{i})^{2}-\frac{k_{1}}{2}x^{2}_{i}
		\label{hall} 
		\end{equation}
		where  $c$  denotes the speed of light in vacuum. The first term of the above Lagrangian (\ref{hall}) represent the kinetic term of the charges and the second term represent their interaction with the external magnetic field $B$. Further, to prescribe a vector potential $\vec{A}$, satisfying $\vec{\nabla}\times \vec{A}=B\hat{z}$, we chose rotationally symmetric gauge. The third term is the harmonic interaction between the two charges and finally the fourth term describes
		additional interactions of the negative charge with an impurity.\\
		Introducing magnetic length scale $l_{B}=\sqrt{\frac{\hbar c}{eB}}$~\cite{PhysRevLett.69.3001,PhysRevB.29.5617} and the dimensionless coordinates $\xi_{i}=l^{-1}_{B}x_{i}$ and $\rho_{i}=l^{-1}_{B}y_{i}$, the Lagrangian (\ref{hall}) reduces to -
		\begin{equation}
		L=\frac{\hbar}{2}\bigg(\frac{mc}{eB}(\dot{\xi}^{2}+\dot{\rho}^{2})+\epsilon_{ij}(\xi_{j}\dot{\xi}_{i}-\rho_{j}\dot{\rho}_{i})-\frac{ck_{0}}{eB}(\xi_{i}-\rho_{i})^{2}-\frac{ck_{1}}{eB}\rho^{2}_{i}\bigg).
		\label{lan}
		\end{equation} 
		In what follows we will be interested in the limit of strong magnetic field $B$ and small mass $m$ as $\frac{m}{eB}\rightarrow 0$, in which the kienetic term from (\ref{lan}) can be effectively ignored \cite{doi:10.1142/S0217751X04018099}. Also, upon quantization of the system (\ref{hall}), in the  absence of a harmonic interaction potential between two charges and the additional interactions ("impurities"), the quantum spectrum consists of the well-known Landau levels \cite{PhysRevD.41.661,PhysRevD.70.107701}. Since the separation between the successive  Landau levels  is $\mathcal{O}(\frac{eB}{m})$, if the magnetic field is
		strong, only the lowest Landau level (LLL) is relevant for the dynamics. The higher states are essentially
		decoupled to infinity. So the
		Lagrangian can be modified so as to describe only high magnetic field effects as mentioned just above by setting $\frac{m}{eB}$
		to zero in (\ref{lan}) \cite{Hellerman_2001, Sourrouille:2020zgf, JACKIW200230}. Thus the Lagrangian of interest here, in terms of the dimensionful Cartesian co-ordinates $x_{i}$ and $y_{i}$, is the following :-
		\begin{equation}
		L_{0}=\frac{eB}{2c}(\epsilon_{ij}x_{j}\dot{x}_{i}-\epsilon_{ij}y_{j}\dot{y}_{i})-V(x_{i},y_{i})~,
		\label{sp}
		\end{equation}
		with $V(x_{i},y_{i})=\frac{k_{0}}{2}(x_{i}-y_{i})^{2}+\frac{k_{1}}{2}x^{2}_{i}$. Thus the Euler-Lagrange equations associated with the above first-order Lagrangian (\ref{sp}) is given by -
		\begin{equation}
		\dot{x_{i}}=\frac{c}{eB} \epsilon_{ij}\frac{\partial V}{\partial x_{j}}~~~; \dot{y_{i}}=-\frac{c}{eB} \epsilon_{ij}\frac{\partial V}{\partial y_{j}}.
		\label{lag}
		\end{equation}
		The Hamiltonian corresponding to (\ref{sp}) is constructed by the usual Legendre transformation :
		\begin{equation}
		H_{0}=\frac{\partial L_{0}}{\partial \dot{x}_{i}}\dot{x}_{i}+\frac{\partial L_{0}}{\partial \dot{y}_{i}}\dot{y}_{i}-L_{0}=V(x_{i},y_{i}).
		\label{S}		
		\end{equation}
		In order to show the equivalence between the Lagrangian
		and Hamiltonian formalisms \cite{Fuji,BANERJEE1999248} , we consider the Hamilton's equations of motion:
		\begin{equation}
		\dot{x_{i}}=\{x_{i},H_{0} \}=\{x_{i},V \} ,
		\label{gg}
		\end{equation}
		\begin{equation}
		\dot{y_{i}}=\{y_{i},H \}=\{y_{i},V \} ,
		\label{h} 
		\end{equation}
		with the potential $V(x_{i},y_{i})$ playing the role of the Hamiltonian.
		The symplectic structure can readily be obtained now by comparing the Lagrangian equations of motion (\ref{lag}) with the form of Hamilton's equations of motion (\ref{gg},\ref{h}) to yield the following brackets :
		\begin{equation}\label{sym1}
		\{x_{i},x_{j}\}=\frac{c}{eB} \epsilon_{ij};~~\{y_{i},y_{j}\}=-\frac{c}{eB} \epsilon_{ij};~\{x_{i},y_{j}\}=0.
		\end{equation} 
		Let us define a new pair of canonical variables as,
		\begin{equation}
		p_{i}=\frac{eB}{c}\epsilon_{ij}(x_{j}-y_{j})~~and ~~x_{i},
		\label{deg}
		\end{equation}
		which satisfies the following symplectic structure :
		\begin{equation}
		\{x_{i},x_{j}\}=\frac{c}{eB} \epsilon_{ij};~~\{x_{i},p_{j}\}= \delta_{ij};~\{p_{i},p_{j}\}=0,
		\label{sm}
		\end{equation}
		where $p_{i}$ play the role of canonical conjugate momentum of cordinates $x_{i}$. Thus at very high magnetic field and low mass limit, the canonical Hamiltonian (\ref{S}) can be rewritten as -
		\begin{equation}
		H_{0}=\frac{p_{i}^2}{2m_{B}}+\frac{1}{2}k_{1}x_{i}^2,
		\label{hamil}
		\end{equation}
		where $ m_{B}=\frac{e^2B^2}{k_{0}}$. Therefore at strong magnetic field limit, the dynamics of system (\ref{hall}) is governed by a two dimensional harmonic oscillator with the deformed symplectic structure (\ref{sm}). Notice that Poisson-noncommutativity of the coordinates has been already established at the classical level as the symplectic bracket between
		coordinates $x_{i}$ is nonvanishing. In the next section we shall discuss the quantum version of the theory. 
		
		\section{ Quantization: The emergence of noncommutative quantum mechanics in
			high magnetic field }  
		Now to describe quantum theory of the above model at stong magnetic field limit in a systematic manner, we start with the hermitian Hamiltonian operator:
		\begin{equation}
		\hat{H}=\frac{\hat{p}_{i}^{2}}{2m_{B}}+ \frac{1}{2} k_{1} \hat{x}_{i}^{2},
		\label{Ham}
		\end{equation}
		where phase-space variables/operators ($ \hat{x}_{i}, \hat{p}_{i}$) satisfy the following noncommutative Heisenberg algebra (NCHA) :
		\begin{equation}
		[\hat{x}_{i},\hat{x}_{j}]=i\theta \epsilon_{ij},~ [\hat{x}_{i},\hat{p}_{j}]=i\hbar\delta_{ij}, ~[\hat{p}_{i},\hat{p}_{j}]=0;~~ for~ i,j=1,2
		\label{hup} 
		\end{equation}
		with $\theta=\frac{\hbar c}{eB}>0$. So, the system of interest (\ref{Ham}) is nothing but a two-dimensional harmonic oscillator system placed in the ambient noncommutative space \cite{PhysRevD.66.027701}, which we refer here as exotic oscillators.\\
		
		Recently in \cite{Scholtz_2007, Scholtz_2009}, it was pointed out that noncommutative quantum mechanics should be formulated as a quantum system on the Hilbert space of Hilbert-Schmidt operators acting on the classical configuration space. Here, we present a very brief review of the formulation in order to dispense it with an appropriate physical setting and also to pave the road for path-integral scheme.  
		In two dimension, the coordinates of configaration space follow the NC algebra :
		\begin{equation}\label{coordinatealg}
		[\hat{x}_{i},\hat{x}_{j}]=i\theta \epsilon_{ij};  ~for~i=j=1,2
		\end{equation}
		and they act on a Hilbert space $\mathcal{H}_{c}$, which is referred as configuration space or classical Hilbert space. One can realize it by introducing the annihilation and creation operators,
		\begin{equation}
		\hat{b}=\frac{1}{\sqrt{2\theta}}(\hat{x}_{1}+i\hat{x}_{2}),~~\hat{b}^{
			\dagger}=\frac{1}{\sqrt{2\theta}}(\hat{x}_{1}-i\hat{x}_{2}),
		\end{equation}
		satisfying the commutation relation $[\hat{b},\hat{b}^{\dagger}]=\mathbb{I}_{c}$. The noncommutative configuration space is isomorphic to boson Fock space :
		\begin{equation}
		\mathcal{H}_{c}=~span\{\left|n\right\rangle= \frac{1}{\sqrt{n!}}(\hat{b}^{\dagger})^{n}\left|0\right\rangle\},
		\label{hc}
		\end{equation}
		This $\mathcal{H}_{c}$ furnishes the representation of the coordinate algebra (\ref{coordinatealg}). Now quantum states of our noncommutative Hamiltonian system (\ref{Ham}) are operators which are elements of the algebra generated by the coordinate operators i.e.$\psi(\hat{x}_{1},\hat{x}_{2})$. These states live in a Hilbert space, called the quantum Hilbert space $\mathcal{H}_{q}$ \cite{Scholtz_2009,Rohwer_2010} in which the entire NCHA (\ref{hup}) should be represented. In fact, one can identify $\mathcal{H}_{q}$ as $\mathcal{H}_{c} \otimes \mathcal{H}^{*}_{c}$, with $\mathcal{H}^{*}_{c}$ being the dual of $\mathcal{H}_{c}$ (\ref{hc}). An important notation that we adopt is the following: states in noncommutative configuration space are denoted by $\left|~\right\rangle$ and states in the quantum Hilbert space are  denoted by a round " bra-ket" $\psi(\hat{x}_{1},\hat{x}_{2})\equiv |\psi) $. The natural choice for the basis of quantum Hilbert space is the space of Hilbert-Schmidt operators -
		\begin{equation}
		\mathcal{H}_{q}=span\{\ \psi(\hat{x}_{1},\hat{x}_{2})=|\psi):\psi(\hat{x}_{1},\hat{x}_{2})\in \mathcal{B}(\mathcal{H}_{c}), tr_{c}(\psi^{\dagger}(\hat{x}_{1},\hat{x}_{2})\psi(\hat{x}_{1},\hat{x}_{2}))< \infty\}\,
		\label{hq}
		\end{equation}\normalsize
		where the subscript $c$ refers to tracing over $\mathcal{H}_{c}$ and $\mathcal{B}(\mathcal{H}_{c})$ is the set of bounded operators acting on $\mathcal{H}_{c}$. Now we search for a unitary representation of the non-commutative  Heisenberg algebra (\ref{hup}) on quantum Hilbert space in terms of operators $\hat{X}_{i}$, $\hat{P}_{i}$ in the following way -
		\begin{equation}\label{qh1}
		\hat{X}_{i} |\psi)=|\hat{x}_{i}\psi),~~ \hat{P}_{i}=\frac{\hbar}{\theta}\epsilon_{ij}|[\hat{x}_{j},\psi])=\frac{\hbar}{\theta}\epsilon_{ij} [\hat{X}_{i}-\hat{X}^{R}_{j}]||\psi).
		\end{equation}  
		Here the capital letters $\hat{X}_{i}$ and $\hat{P}_{i}$  have been used to denote the  representations of the operators $\hat{x}_{i}$ and $\hat{p}_{i}$ acting on $\mathcal{H}_{q}$. Note that we
		have taken the action of $\hat{X}_{i}$ to be left action
		by default and the momentum operator acts
		adjointly. Also, $\hat{X}^{R}_{i}$ implies the right action on the quantum Hilbert space in the following way :
		\begin{equation}\label{qh2}
		\hat{X}^{R}_{i}|\psi)=|\psi)\hat{X}_{i}=|\psi \hat{x}_{i});~ [\hat{X}^{R}_{i},\hat{X}^{R}_{j}]|\psi)=|\psi[\hat{x}_{j},\hat{x}_{i}])=-\theta \epsilon_{ij}|\psi)~~\forall \psi \in \mathcal{H}_{q}
		\end{equation}
		It can be checked easily that the right action satisfies
		\begin{equation}
		[\hat{X}_{i},\hat{X}^{R}_{j}]=0
		\end{equation}
		We can thus identify $\hat{X}^{R}_{i}$ as quantum mechanical counterpart of $y_{i}$ in (\ref{deg}) and note (\ref{qh1}, \ref{qh2}) are analogous to (\ref{sym1}, \ref{deg}) in the quantum setting. Therefore, the classical setting discussed in previous section provides a clear physical interpretation of the abstract formulation of noncommutative mechanics in terms of Hilbert-Schmidt operators. Having pointed this, we now introduce one further notational convention - for any operator $ \hat{O}$  acting on the quantum Hilbert space (\ref{hq}), we may
		define left and right action (denoted by superscripted L and R) as follows:
		\begin{equation}
		\hat{O}^{L}|\psi)=\hat{O}|\psi),~\hat{O}^{R}|\psi)=|\psi)\hat{O};~~\forall |\psi)=\psi(\hat{x}_{1},\hat{x}_{2})\in \mathcal{H}_{q}
		\end{equation}
		Thus the operators acting on the quantum Hilbert space obey the commutation relations -
		\begin{equation}
		[\hat{X}_{i},\hat{X}_{j}]=i\theta \epsilon_{ij},~[\hat{X}_{i},\hat{P}_{j}]=i\hbar\delta_{ij},~[\hat{P}_{i},\hat{P}_{j}]=0~~ for~ i,j=1,2;
		\label{bd}
		\end{equation} 
	It is now useful to define the following operators on the quantum Hilbert space which we shall require later:
	\begin{equation}
		\hat{B}=\frac{\hat{X}_{1}+i\hat{X}_{2}}{\sqrt{2\theta}},~ \hat{B}^{\ddagger}=\frac{\hat{X}_{1}-i\hat{X}_{2}}{\sqrt{2\theta}},~\hat{P}=\hat{P}_{1}+i\hat{P}_{2},~ \hat{P}^{\ddagger}=\hat{P}_{1}-i\hat{P}_{2};~~ [\hat{B},\hat{B}^{\ddagger}]=\mathbb{I}_{q}
		\label{cx}
	\end{equation} 
	where we have used the symbol $\ddagger$ specifically for the operator adjoint on $\mathcal{H}_{q}$. It can be easily checked that all these phase space operators are self adjoint with respect to the inner product -
	\begin{equation}
		(\psi|\phi)=tr_{c}(\psi^{\dagger}\phi)< \infty  ~~\forall \psi, \phi \in \mathcal{H}_{q} ~,
	\end{equation}
	and stems from the fact that the product of two HS operators has finite trace-class norm. 
		Therefore, the representation of our system Hamiltonian (\ref{Ham}) on $\mathcal{H}_{q}$ is given by
		\begin{equation}
		\hat{H}=\frac{1}{2m_{B}}(\hat{P}_{1}^{2}+\hat{P}_{2}^{2})+ \frac{1}{2} k_{1} (\hat{X}_{1}^{2}+\hat{X}_{2}^{2}),
		\label{ham}
		\end{equation}
		where phase-space operators $\hat{X}_{i} $ and $\hat{P}_{i}$ satisfy the same i.e. isomorphic algebra (\ref{bd})
		while acting on the quantum Hilbert space $\mathcal{H}_{q}.$ Here we do not embark upon the canonical quantization of (\ref{ham}) as this has been studied extensively in the literature \cite{Rohwer_2010,Kumar:2013gca} where the system gets diagonalized as two decoupled oscillators of frequency $\omega_{\pm}$ :
	\begin{equation}
		\hat{H}\left|n_{1},n_{2}\right)=E_{n_{1}n_{2}}\left|n_{1},n_{2}\right);~~E_{n_{1},n_{2}}= (n_{1}\hbar\omega_{+}+n_{2}\hbar\omega_{-})+E_{0,0}
		\label{drf}
	\end{equation}
	where the characteristic frequencies $\omega_{\pm}$ are given by, $ \omega_{\pm}=\omega\sqrt{1+\frac{m_B^2 \omega^2 \theta^2}{4\hbar^2}} \pm \frac{m_{B}\omega^{2}\theta}{2\hbar}$ and  $E_{0,0}=\frac{1}{2}\hbar (\omega_{+}+\omega_{-})$ is the finite zero point energy of the system Hamiltonian. This however, being a $c$-number and not an operator will yield the same value in all states and is displaced off by simply defining the zero of energy as,
		\begin{equation}
		\hat{\widetilde{H}}:=\hat{H}- E_{0,0}\mathbb{I}_{q} ,
		\label{nh}
		\end{equation}
		Therefore 
		\begin{equation}
		\hat{\tilde{H}}\left|n_{1},n_{2}\right) =\tilde{E}_{n_{1}n_{2}}\left|n_{1},n_{2}\right),
		\label{nope}
		\end{equation}
		with $\hat{\widetilde{H}}|0,0)=0$ and $\tilde{E}_{n_{1}n_{2}}=n_{1}\hbar\omega_{+}+n_{2}\hbar\omega_{-}$. From now on, we only work with $\hat{\widetilde{H}}.$ As is the case always, the energy eigenvectors of $\mathcal{H}_{q}$ constitute a complete set-
		\begin{equation}
			\sum_{n_{1},n_{2}}|n_{1},n_{2})(n_{1},n_{2}|=\mathbb{I}_{q}
		\end{equation}
		We now, as discussed at the outset, proceed to carry out the path-integral of the system in order to study its scaling properties. 
		
\section{Coherent state and the path integral}
		We would like to introduce the notion of position states, however, in view of the absence of common eigenstates of
		$\hat{x}_{1} $ and $ \hat{x}_{2}$, the best one can do is to introduce the minimal uncertainty states i.e. maximally
		localized states (coherent states) on $\mathcal{H}_{c}$ \cite{Klauder} as,
		\begin{equation}
		\left|z \right\rangle=e^{-\bar{z}\hat{b}+z \hat{b}^{\dagger}} \left|0 \right\rangle= e^{-\frac{\mid z\mid^{2}}{2}} e^{z\hat{b}^{\dagger}}\left|0 \right\rangle ~\in \mathcal{H}_{c},
		\end{equation}
		where $z=\frac{x_{1}+i x_{2}}{\sqrt{2\theta}}$ is a dimensionless complex number. These states provide an over-complete basis on $\mathcal{H}_{c}$ and it is possible to write the identity as,
		\begin{equation}
		\int \frac{dz d\bar{z}}{\pi}  \left|z \right\rangle \langle z|=\mathbb{I}_{c},
		\end{equation}
		 Using these states, we can construct a state in quantum Hilbert space $\mathcal{H}_{q}$ as follows :
		\begin{equation}
		|z,\bar{z})=|x_{1},x_{2})=\frac{1}{\sqrt{2\pi\theta}}\left|z \right\rangle \langle z|~\in \mathcal{H}_{q};~~~(w.\bar{w}|z,\bar{z})=\frac{1}{2\pi\theta}e^{-\mid w-z\mid^{2}},
		\end{equation}
		satisfying  $\hat{B}|z,\bar{z})=z|z,\bar{z})$. Now, the coherent state ("position") representation of a state $|\psi)=\psi(\hat{x}_{1},\hat{x}_{2})$ can be expressed as,
		\begin{equation}
		\psi(x_{1},x_{2})=(z,\bar{z}|\psi)=\frac{1}{\sqrt{2\pi\theta}}tr_{c}( \left|z \right\rangle \langle z|\psi(\hat{x}_{1},\hat{x}_{2})) = \frac{1}{\sqrt{2\pi\theta}}\langle z|\psi(\hat{x}_{1},\hat{x}_{2})\left|z \right\rangle.
		\end{equation}
		We now introduce the normalized momentum eigenstates \cite{PhysRevLett.102.241602}  such that -
		\begin{equation}
		|p,\bar{p})=\sqrt{\frac{\theta}{2\pi\hbar^{2}}} e^{i\sqrt{\frac{\theta}{2\hbar^{2}}}(\bar{p}\hat{b}+p\hat{b}^{\dagger})};~\hat{P}_{i}|p,\bar{p})=p_{i}|p,\bar{p}),~ p=p_{1}+ip_{2}
		\end{equation}
		with these states satisfying usual resolution of identity and orthogonality conditions, one has the coherent state representation of the momentum states on the noncommutative plane to be given by \cite{PhysRevLett.102.241602}-
		\begin{equation}
		(z,\bar{z}|p,\bar{p})=\frac{1}{\sqrt{2\pi\hbar^{2}}}e^{-\frac{\theta}{4\hbar^{2}}p\bar{p}} e^{i\sqrt{\frac{\theta}{2\hbar^{2}}}(p\bar{z}+\bar{p}z)}.
		\end{equation}
		Now the completeness relations for the "position basis" (coherent state basis) $|z,\bar{z})$ on $\mathcal{H}_{q}$,  given by  Scholtz et al.\cite{Scholtz_2009} reads,
		\begin{equation}
		\int 2\theta dzd\bar{z} ~|z,\bar{z})\star_{V} (z,\bar{z}|= \int dx_{1}dx_{2} ~|x_{1},x_{2})\star_{V} (x_{1},x_{2}|= \mathbb{I}_{q}.
		\label{cor}
		\end{equation}
		This involves the Voros star product \cite{PhysRevA.40.6814}, where the star-product between two functions $f(z,\bar{z})$ and $g(z,\bar{z})$ is defined as-
		\begin{equation}
		f(z,\bar{z})\star_{V} g(z,\bar{z})=f(z,\bar{z}) e^{\stackrel{\leftarrow}{\partial_{\bar{z}}}
			\stackrel{\rightarrow}{\partial_z}}  g(z,\bar{z}) 
		\end{equation}
		
		In fact, in \cite{PhysRevD.70.105007,PhysRevD.72.065014}, it has been pointed out that the low energy dynamics of
		relativistic quantum field theories in the presence of strong magnetic field can be described by the Voros star product. The finite time propagation kernel can be splitted up into n steps,
		\footnotesize
		\begin{equation}
		(z_{f},t_{f}|z_{0},t_{0})=\lim_{n\to\infty}\int^{\infty}_{-\infty} (2\theta)^{n}\prod_{j=1}^{n}  dz_{j}d\bar{z}_{j}(z_{f},t_{f}|z_{n},t_{n})\star_{V_n}(z_{n},t_{n}|...........|z_{1},t_{1})\star_{V_{1}}(z_{1},t_{1}|z_{0},t_{0})
		\label{ker}
		\end{equation}
		\normalsize
		Writing down the propagation over an infinitesimal  step $\tau=\frac{t_{f}-t_{0}}{n+1}:$ 
		\small
		\begin{eqnarray}
		(z_{j+1}, t_{j+1}|z_j, t_j)&=&(z_{j+1}|e^{-\frac{i}{\hbar}\tau\hat{\widetilde{H}}}|z_j)\nonumber\\
		&=&\int_{-\infty}^{+\infty}d^{2}p_j~e^{-\frac{\theta}{2\hbar^{2}}\bar{p}_j p_{j}}
		e^{i\sqrt{\frac{\theta}{2\hbar^{2}}}\left[p_{j}(\bar{z}_{j+1}-\bar{z}_{j})+\bar{p}_{j}(z_{j+1}-z_{j})\right]}\nonumber\\
		&&~~~~~~~~~~~~~~~~~~~~~~~~~~~~~~\times e^{-\frac{i}{\hbar}\epsilon[\frac{\bar{p}_j p_{j}}{2m_{B}}+\frac{k_{1} \theta}{2}(2\bar{z}_{j+1}z_{j}+c_{0})]}
		\label{inftm}
		\end{eqnarray}
	\normalsize
		where $c_{0}=1-\frac{2}{\theta k_{1}} E_{0,0}$ . Now substituting the above expression in (\ref{ker}) and computing star products explicitly, we obtain (apart from a constant factor) - 
	\small
	\begin{equation}
		\begin{alignedat}{1}
	& (z_f, t_f|z_0, t_0)\\ & =e^{-\vec{\partial}_{z_f}\vec{\partial}_{\bar{z}_0}}\lim_{n\rightarrow\infty}\int \prod_{j=1}^{n} (dz_{j}d\bar{z}_{j})
	\prod_{j=0}^{n}d^{2}p_{j}  \exp\sum_{j=0}^{n}\tau\left[\frac{i}{\hbar}\sqrt{\frac{\theta}{2}}\left[p_{j}\left\{\frac{\bar{z}_{j+1}-\bar{z}_{j}}{\tau}\right\}+\bar{p}_{j}\left\{\frac{z_{j+1}-z_{j}}{\tau}\right\}-\sqrt{2\theta} k_{1}\bar{z}_{j+1}z_{j}\right]\right]\\
	&\exp\sum_{j=0}^{n}\tau\left[-\frac{i}{2m\hbar} \bar{p}_{j}p_{j}+\frac{\theta}{2\hbar^2}\frac{(p_{j+1}-p_{j})}{\tau}\bar{p}_{j}\right],
		\end{alignedat}
		\label{ax}
	\end{equation}
\normalsize
		 where, $\sigma=-i(\frac{\tau}{2m_{B}\hbar}-i\frac{\theta}{2\hbar^{2}})$ and we treat $z_{n+1}=z_{f}$. Finally, in the $\epsilon\rightarrow0$ or $n\rightarrow \infty $ limit, symbolically, we write the phase space form of the path integral:
		\small
	    \begin{equation}
		(z_f, t_f|z_0, t_0)=e^{-\vec{\partial}_{z_f}\vec{\partial}_{\bar{z}_0}}\int^{z(t_{f})=z_{f}}_{z(t_{0})=z_{0}} \mathcal{D}\bar{z}(t)\mathcal{D} z(t)\int \mathcal{D}\bar{p}(t)\mathcal{D} p(t)e^{\frac{i}{\hbar}S_0 [z(t),\bar{z}(t),p(t),\bar{p}(t)]}
		\end{equation}
	\normalsize
		where $S_0$ is the phase-space action given by -
		\begin{equation}
		S_0=\int^{t_{f}}_{t_{0}} dt \left[ \sqrt{\frac{\theta}{2}} (p\dot{\bar{z}}+\bar{p}\dot{z})- \frac{i\theta}{2\hbar}\bar{p}\dot{p}-\frac{1}{2m_{B}}\bar{p}p-\theta k_{1}\bar{z}z \right].
		\label{action}
		\end{equation}
		Note that this action contains an exotic $\bar{p}\dot{p}$ term which is a Chern-Simons like term in momentum space and is responsible for rendering the configuration space action to be non-local \cite{Gangopadhyay_2014}. On recognizing that the first order system (\ref{action}) can be described in the extended phase
		space, where $z(t)$ and $p(t)$ are treated as configuration space variables of the system, the action can be written in the so-called symplectic form \cite{PhysRevLett.60.1692} as -
		\begin{equation}
		S_0=\int^{t_{f}}_{t_{i}}dt \left[ \frac{1}{2}\xi^{\alpha} f_{\alpha \beta}\dot{\xi^{\beta}}-V(\xi)\right] ,~~~~\xi^{\alpha}:=\{z, \bar{z}, p,\bar{p}\};~~~\alpha=1,2..,4
		\end{equation}   
		with  
		\begin{equation}
		f = \begin{pmatrix}
		\huge0 & -\sqrt{\frac{\theta}{2}}\sigma_{1}   \\
		\sqrt{\frac{\theta}{2}}\sigma_{1} & -\frac{\theta}{2}\sigma_{2}
		\end{pmatrix}.
		\end{equation} 
		where $\sigma_{1}$, $\sigma_{2}$ are the usual Pauli matrices, and 
		\begin{equation}
		V(\xi)=\frac{1}{2m_{B}}\bar{p}p+\theta k_{1}\bar{z}z 
		\end{equation}
		One can easily read off the symplectic brackets :-
		\begin{equation}
		\{z,\bar{z}\}=-\frac{i}{\hbar},~~~~~\{z,\bar{p}\}=\{\bar{z},p\}=\sqrt{\frac{2}{\theta}},~~~~ \{p,\bar{p}\}=0
		\label{symplectic}
		\end{equation}
		The symplectic brackets are consistent with the classical version
		of the noncommutative Heisenberg algebra  given in (\ref{sm}). Having done this brief review to describe the path-integral action of the noncommutative harmonic oscillator, we are all set to dive into the main theme of the paper where now we will build-up on this action to study its scaling properties.\\

		Looking at the expression of the path integral kernel, we can think of noncommutative quantum mechanics as a complex scalar field theory in $0+1$ dimensions, the coordinates are field which depend on time $z(t)$. Therefore we write the matrix elements of the
		time-ordered product of source term for all $t$ with $t_{i}<t<t_{f}$ as a path integral -
		\small
			\begin{equation}
			\begin{alignedat}{1} (z_{f},t_{f}|z_{0},t_{0})_{J,\bar{J}} & =(z_{f},t_{f}|\large{T}exp\left[\frac{i}{\hbar}\int^{t_{f}}_{t_{i}} \sqrt{\frac{\theta}{2}}(J(t)\hat{B}^{\dagger}+\bar{J}(t)\hat{B}) dt  \right]|z_{0},t_{0})\\
			& =e^{-\vec{\partial}_{z_f}\vec{\partial}_{\bar{z}_0}}\int^{z(t_{f})=z_{f}}_{z(t_{0})=z_{0}}\mathcal{D}\bar{z}\mathcal{D} z \mathcal{D}\bar{p}\mathcal{D} p~~ exp\left[\frac{i}{\hbar}(S_0+\int^{t_{f}}_{t_{i}}\sqrt{\frac{\theta}{2}}[\bar{z}(t)J(t)+z(t)\bar{J}(t)]\right]\\
			&
			=e^{-\vec{\partial}_{z_f}\vec{\partial}_{\bar{z}_0}}\int^{z(t_{f})=z_{f}}_{z(t_{0})=z_{0}}\mathcal{D}\bar{z}\mathcal{D} z \mathcal{D}\bar{p}\mathcal{D} p~~ exp\left(\frac{i}{\hbar}S_{J,\bar{J}}\right),
			\end{alignedat}
			\label{oci}
			\end{equation}
		\normalsize
		with 
		\begin{equation}
		S_{J,\bar{J}}= \int ^{t_{f}}_{t_{0}} dt \left[ \sqrt{\frac{\theta}{2}} (p\dot{\bar{z}}+\bar{p}\dot{z})- \frac{i\theta}{2\hbar}\bar{p}\dot{p}-\frac{1}{2m_{B}}\bar{p}p-\theta k_{1}\bar{z}z  +\sqrt{\frac{\theta}{2}}[\bar{z}(t)J(t)+z(t)\bar{J}(t)\right],
		\end{equation}
		where $J(t)=J_{1}(t)+iJ_{2}(t)$  is a time dependent source vanishing  at $ t \rightarrow \pm\infty$.
		Now in quantum theories, the object of prime usefulness is the vacuum to 
		vacuum persistence amplitude in the presence of an external source. A simple way to obtain this is to go back to the transition amplitude
		in the coordinate space (\ref{oci}) and introduce complete sets of energy as :-
		\small
			\begin{equation}
			\begin{alignedat}{1}
			& (z_{f},t_{f}|z_{0},t_{0})_{J,\bar{J}} \\
			& =\sum_{n_{1},n_{2}}\sum_{m_{1},m_{2}}(z_{f},t_{f}|n_{1},n_{2}) (n_{1},n_{2}|\large{T}\exp\left[\frac{i}{\hbar}\int^{t_{f}}_{t_{0}} \sqrt{\frac{\theta}{2}}(J(t)\hat{B}^{\dagger}+\bar{J}(t)\hat{B}) dt  \right]|m_{1},m_{2})\times\\
			& (m_{1},m_{2}|z_{0},t_{0})\\
			& =\sum_{n_{1},n_{2}}\sum_{m_{1},m_{2}}(n_{1},n_{2}|\large{T}\exp\left[\frac{i}{\hbar}\int^{t_{f}}_{t_{0}} \sqrt{\frac{\theta}{2}}(J(t)\hat{B}^{\dagger}+\bar{J}(t)\hat{B}) dt  \right]|m_{1},m_{2})\\
			&~~~~~~~~~~~~~~~\times(m_{1},m_{2}|e^{\frac{i}{\hbar}t_{0}\hat{\tilde{H}}}|z_{0}) (z_{f}|e^{-\frac{i}{\hbar}t_{f}\hat{\tilde{H}}}|n_{1},n_{2})\\
			&
			=(0,0|\large{T}\exp\left[\frac{i}{\hbar}\int^{t_{f}}_{t_{0}} \sqrt{\frac{\theta}{2}}(J(t)\hat{B}^{\dagger}+\bar{J}(t)\hat{B}) dt  \right]|0,0) \times(0,0|z_{0}) (z_{f}|0,0)\\
			&~~~~~~~+\sum_{n_{1},n_{2}\neq 0}\sum_{m_{1},m_{2}\neq 0}(n_{1},n_{2}|\large{T}\exp\left[\frac{i}{\hbar}\int^{t_{f}}_{t_{i}} \sqrt{\frac{\theta}{2}}(J(t)\hat{B}^{\dagger}+\bar{J}(t)\hat{B}) dt  \right]|m_{1},m_{2})\\
			&~~~~~~~~~~~~~~~\times(m_{1},m_{2}|z_{0}) (z_{f}|n_{1},n_{2}) e^{-\frac{i}{\hbar}t_{f}\tilde{E}_{n_{1},n_{2}}+\frac{i}{\hbar}t_{0}\tilde{E}_{m_{1},m_{2}}}~~~,  
			\end{alignedat}
			\label{oct}
			\end{equation}
		\normalsize
	    where we have used the fact of (\ref{nope}). To accomplish the projection onto the vacuum, we now introduce a variable $T$ with units of time. One can for instant replace -
		\begin{equation}
		t_{f}=T(1-i\epsilon), ~~~~~~~t_{0}=-T(1-i\epsilon),
		\end{equation}
		and take $T\rightarrow \infty$, with $\epsilon$ being a small positive constant. At the end  we will take the limit $\epsilon \rightarrow 0$, but only after the infinite limit of $T$ is carried out. In the limit $T\rightarrow \infty$, the exponentials in (\ref{oct}) oscillate out to zero except for the ground
		state. Thus in this asymptotic limit, we obtain
		\small
		\begin{equation}
			\begin{alignedat}{1}
			\lim_{\epsilon\to 0}\lim_{\substack{t_{f}\rightarrow\infty(1-i\epsilon)\\t_{0}\rightarrow-\infty(1-i\epsilon)}}(z_{f},t_{f}|z_{0},t_{0})_{J,\bar{J}}&
			=(0,0|\large{T}exp\left[\frac{i}{\hbar}\int^{\infty}_{-\infty} \sqrt{\frac{\theta}{2}}(J(t)\hat{B}^{\dagger}+\bar{J}(t)\hat{B}) dt  \right]|0,0) \\
			&~~~~~~~~~~~~~~~~~~\times(0,0|z_{0}) (z_{f}|0,0).
			\end{alignedat}
			\end{equation}
		\normalsize
		However, one can also observe that -
		\begin{equation}
		\lim_{\epsilon\to 0}\lim_{\substack{t_{f}\rightarrow\infty(1-i\epsilon)\\t_{0}\rightarrow-\infty(1-i\epsilon)}}(z_{f},t_{f}|z_{0},t_{0})_{J=\bar{J}=0}=(0,0|z_{0}) (z_{f}|0,0)
		\end{equation}
		Consequently, we can write -
		\small
		\begin{equation}
		(0,0|T\exp\left[\frac{i}{\hbar}\int^{\infty}_{-\infty} \sqrt{\frac{\theta}{2}}(J(t)\hat{B}^{\dagger}+\bar{J}(t)\hat{B})dt  \right]|0,0) =\lim_{\epsilon\to 0}\lim_{\substack{t_{f}\rightarrow\infty(1-i\epsilon)\\t_{0}\rightarrow-\infty(1-i\epsilon)}}\frac{(z_{f},t_{f}|z_{0},t_{0})_{J,\bar{J}}}{(z_{f},t_{f}|z_{0},t_{0})_{J=\bar{J}=0}}
		\label{b}
		\end{equation} 
	\normalsize
		As the left hand side of (\ref{b}) is independent of the boundary conditions $z_{0} $ and $z_{f}$ imposed at $t\rightarrow\pm\infty$,
		so the right hand side is also independent of the boundary conditions imposed at $t\rightarrow\pm\infty$
		provided that one chooses the same boundary conditions in the numerator and the
		denominator. Furthermore, the right hand side has the structure of a functional integral
		and we can write (\ref{b}) also as,
		\small
			\begin{equation}
			\begin{alignedat}{1}(0,0|\large{T}exp\left[\frac{i}{\hbar}\int^{\infty}_{-\infty} \sqrt{\frac{\theta}{2}}(J(t)\hat{B}^{\dagger}+\bar{J}(t)\hat{B}) dt  \right]|0,0)&
			=(0,0;\infty|0,0;-\infty)_{J,\bar{J}}\\
			=N^{-1} \int \mathcal{D}\bar{z}\mathcal{D} z \mathcal{D}\bar{p}\mathcal{D} p~~ e^{\frac{i}{\hbar}\left[ S_{0}+\int^{t_{f}=\infty}_{t_{0}=-\infty}\sqrt{\frac{\theta}{2}}[\bar{z}(t)J(t)+z(t)\bar{J}(t)dt ]\right]}, 
			\label{a}
			\end{alignedat}
			\end{equation}
		\normalsize
		where the normalization factor $N$ in (\ref{a}) is fixed so as to ensure $(0,0|0,0)=1$.\\
		
		Therefore, the generating functional for connected Green's functions or the vacuum persistence amplitude in
		presence of external source $J(t)$ and $\bar{J}(t)$ is defined as,
		\begin{equation}
		Z(J,\bar{J})=Z(J_{1},J_{2})= N^{-1} \int \mathcal{D}\bar{z}\mathcal{D} z \mathcal{D}\bar{p}\mathcal{D} p~~ e^{\frac{i}{\hbar}\left[ S_{0}+\int^{t_{f}=\infty}_{t_{0}=-\infty}\sqrt{\frac{\theta}{2}}[\bar{z}(t)J(t)+z(t)\bar{J}(t)]dt \right]}.
		\label{me} 
		\end{equation}
		Now, notice that we can realize the symplectic  brackets (\ref{symplectic}) in terms of canonical variables as,   
		\begin{equation}
		\begin{alignedat}{1}z(t) & =\alpha(t)+\frac{i}{2\hbar}\sqrt{\frac{\theta}{2}}p(t)\\
		\bar{z}(t) & =\bar{\alpha}(t)-\frac{i}{2\hbar}\sqrt{\frac{\theta}{2}}\bar{p}(t),
		\end{alignedat}
		\label{bps2}
		\end{equation}
		satisfying 
		\begin{equation}
		\{\alpha,\bar{\alpha}\}=0,~~~~~\{\alpha,\bar{p}\}=\{\bar{\alpha},p\}=\sqrt{\frac{2}{\theta}},~~~~ \{p,\bar{p}\}=0
		\end{equation}
	In order to compute the generating functional in configuration space in a local form, it will be beneficial at this stage to use this change of variables (\ref{bps2})  (this leaves the integration measure invariant) in the noncommutative phase-space generating functional (\ref{me}) to re-write it as :
		\small
		\begin{equation}
		Z(J,\bar{J})=Z(J_{1},J_{2})= N^{-1} \int \mathcal{D}\bar{\alpha}\mathcal{D} \alpha \mathcal{D}\bar{p}\mathcal{D} p~ \exp(\frac{i}{\hbar}S_{J,\bar{J}} [\alpha,\bar{\alpha},p,\bar{p}]),
		\label{le} 
		\end{equation}
		where
		\begin{eqnarray}
		\label{nctrans}
		S_{J,\bar{J}} [\alpha,\bar{\alpha},p,\bar{p}]=\int_{-\infty}^{\infty}dt &&\!\!\!\!\!\!\!\!\!\  \left[\sqrt{\frac{\theta}{2}} (\dot{\bar{\alpha}}-\frac{i\theta}{2\hbar}k_{1}\bar{\alpha}+\frac{i}{\hbar}\sqrt{\frac{\theta}{2}}\bar{J})p+ \sqrt{\frac{\theta}{2}} (\dot{\alpha}+\frac{i\theta}{2\hbar}k_{1}\alpha-\frac{i}{\hbar}\sqrt{\frac{\theta}{2}}J)\bar{p}\right. \nonumber\\
		&& \left. -\frac{1}{2m_{B}}(1+\frac{k_{1}\theta^{2}m_{B}}{4\hbar^{2}})\bar{p}p-\theta k_{1}\bar{z}z +\sqrt{\frac{\theta}{2}}(\alpha \bar{J}+\bar{\alpha}J)\right]
		\end{eqnarray}
	\normalsize
		Furthermore for simplicity, on rescaling the dynamical variables as-
		\begin{equation}
		\begin{alignedat}{1}\alpha(t)\rightarrow Z(t)  & = \frac{1}{\sqrt{1+\frac{k_{1}\theta^{2}m_{B}}{4\hbar^{2}}}}\alpha(t)\\
		p(t)\rightarrow P(t) & =\sqrt{1+\frac{k_{1}\theta^{2}m_{B}}{4\hbar^{2}}}p(t)~,
		\end{alignedat}
		\label{bps1}
		\end{equation}
		the functional integral (\ref{le}) yields :-
		\begin{equation}
		Z(J,\bar{J})=Z(J_{1},J_{2})= N^{-1} \int \mathcal{D}\bar{Z}\mathcal{D} Z \mathcal{D}\bar{P}\mathcal{D} P~e^{\frac{i}{\hbar}S_{J,\bar{J}} [Z,\bar{Z},P,\bar{P}]}
		\label{dge} 
		\end{equation}
		with
		\small
		\begin{eqnarray}
		\label{nct}
		S_{J,\bar{J}} [Z,\bar{Z},P,\bar{P}]&&=\int_{-\infty}^{\infty}dt\left[\sqrt{\frac{\theta}{2}} (\dot{\bar{Z}}-\frac{i\theta}{2\hbar}k_{1}\bar{Z}+\frac{i\lambda}{2\hbar}\sqrt{\frac{\theta}{2}}\bar{J})P+ \sqrt{\frac{\theta}{2}} (\dot{Z}+\frac{i\theta}{2\hbar}k_{1}Z-\frac{i\lambda}{2\hbar}\sqrt{\frac{\theta}{2}}J)\bar{P}\right. \nonumber\\
		&& \left. -\frac{1}{2m_{B}}\bar{P}P-\theta k_{1}(1+\frac{k_{1}\theta^{2}m_{B}}{4\hbar^{2}})\bar{Z}Z+\lambda(1+\frac{k_{1}\theta^{2}m_{B}}{4\hbar^{2}})\sqrt{\frac{\theta}{2}}(Z \bar{J}+\bar{Z}J)\right]
		\label{dbi}
		\end{eqnarray}
		\normalsize
		where $\lambda=(1+\frac{k_{1}\theta^{2}m_{B}}{4\hbar^{2}})^{-\frac{1}{2}}.$  The generating function in configuration space is now easily derived by integrating over the momenta. Indeed, the dependence on momenta in the exponent of (\ref{dge}) is at most quadratic and one may perform the gaussian integrations over the momenta to obtain -
		\begin{equation}
		Z(J,\bar{J})=Z(J_{1},J_{2})= \tilde{N}^{-1} \int \mathcal{D}\bar{Z}\mathcal{D} Z ~ e^{\frac{i}{\hbar}S_{eff} [Z,\bar{Z}]} e^{\frac{i}{\hbar}S_I[Z,\bar{Z},J,\bar{J}]},
		\label{magt}
		\end{equation}
		where $\tilde{N}$ is a normalization constant which arises from the integration of momenta, and 
		\begin{eqnarray}
		\label{nct0}
		S_{eff} [Z,\bar{Z}]=\int_{-\infty}^{\infty}dt &&\!\!\!\!\!\!\!\!\!\  \left[m_{B}\theta \dot{\bar{Z}}\dot{Z} -\frac{im_{B}\theta^{2}k_{1}}{2\hbar}(\bar{Z}\dot{Z}-\dot{\bar{Z}}Z)-k_{1}\theta \bar{Z}Z\right],
		\end{eqnarray}
			\begin{eqnarray}
		\label{ncte}
		S_I[Z,\bar{Z},J,\bar{J}]=\int_{-\infty}^{\infty}dt\lambda\sqrt{\frac{\theta}{2}}    \left[\bar{Z}(J+\frac{i m_{B}\theta}{2\hbar}\dot{J})+Z(\bar{J}-\frac{i m_{B}\theta}{2\hbar}\dot{\bar{J}})+\frac{m_{B}\lambda\theta}{2\hbar^{2}}\sqrt{\frac{\theta}{2}}\bar{J}J\right]
		\end{eqnarray}
		By introducing a "renormalized" (modified) source (and its corresponding complex conjugate)-
		\begin{equation}
		J_{R}(t)=(1+\frac{im_{B}\theta}{2\hbar}\partial_{t})J(t),
		\end{equation}
		and after ignoring the boundary contribution, the above interacting part of the action (\ref{ncte}) can be casted into the standard form -
		\begin{eqnarray}
		\label{octe}
		S_I [Z,\bar{Z},J_{R},\bar{J}_{R}]=\int_{-\infty}^{\infty}dt\lambda\sqrt{\frac{\theta}{2}} \left[\bar{Z}J_{R}+Z\bar{J}_{R}+\frac{m_{B}\lambda\theta}{2\hbar^{2}}\sqrt{\frac{\theta}{2}}\bar{J}_{R}(1+\frac{im_{B}\theta}{2\hbar}\partial_{t})^{-2}J_{R}\right]
		\end{eqnarray} 
		Here the interaction part of the action (\ref{octe}) contains standard source terms as well as a new quadratic source term. Now writing, $Z(t)=\frac{q_{1}(t)+iq_{2}(t)}{\sqrt{2\theta}}$, $J_{R}(t)=J^{(1)}_{R}+iJ^{(2)}_{R}$, the functional integral (\ref{magt}) becomes with the irrelevant source-independent normalization factor $\tilde{N}^{-1}$ being ignored,
		\begin{equation}
		Z\left[J_{R},\bar{J}_{R}\right]=e^{\frac{im_{B}}{\hbar}(\frac{\lambda \theta}{2\hbar})^{2}\int_{-\infty}^{\infty}dt  \left[\bar{J}_{R}(1+\frac{im_{B}\theta}{2\hbar}\partial_{t})^{-2}J_{R}\right]}Z_{eff}\left[J^{(1)}_{R},J^{(2)}_{R}\right]
		\label{connec}
		\end{equation}
		where
		\begin{equation}
		Z_{eff}\left[J^{(1)}_{R},J^{(2)}_{R}\right]=\int\mathcal{D}q_{1}(t)\mathcal{D}q_{2}(t) ~~~e^{\frac{i}{\hbar}\int^{\infty}_{-\infty} dt\left[ L_{eff}+\lambda(q_{1}J^{(1)}_{R}+q_{2}J^{(2)}_{R})\right]}
		\end{equation}
		with
		\begin{equation}
		L_{eff}=\frac{1}{2}m_{B}\dot{q}^{2}_{i}-\frac{m_{B}\theta k_{1}}{2\hbar}\epsilon_{ij}q_{j}\dot{q}_{i}-\frac{1}{2}k_{1} q^{2}_{i}
		\label{effctive}
		\end{equation}
		
		Therefore at strong magnetic field, the classical dynamics of our system (\ref{Ham}) can be described by the effective equivalent Lagrangian (\ref{effctive}), which can be interpreted as the Lagrangian of a charged particle moving in the commutative $q_{1}$-$ q_{2}$ plane under the influence
		of a constant effective magnetic field with an additional quadratic potential where the second term in the right hand side of (\ref{effctive}) describes the interaction of an electrically
		charged particle (of charge e) with a constant magnetic field $(B_{eff})$  pointing along the normal to the plane. The components of the
		corresponding vector potential $\vec{A}_{eff}$ in symmetric gauge can be read off from (\ref{effctive}), 
		\begin{equation}
		(\vec{A}_{eff})_{i}=-\frac{m_{B}\theta k_{1}c}{2 \hbar e} \epsilon_{ij}q_{j},
		\end{equation}
		with $\vec{B}_{eff} = \vec{\nabla}\times \vec{A}_{eff}$. Thus, we have an exact mapping between the planar noncommutative system and (commutative) generalized Landau problem \cite{doi:10.1142/S0217732306019037,PhysRevD.90.047702}, providing an effective commutative description of the noncommutative system. Therefore, the classical dynamics of the noncommutative system is effectively described by the following action :
		\begin{equation}
		S_{eff}=\int dt \left[\frac{1}{2}m_{B}\dot{q}^{2}_{i}+\frac{e}{c}(A_{eff})_{i}\dot{q}_{i}-\frac{1}{2}k_{1} q^{2}_{i}\right].
		\label{nwe ac}
		\end{equation}\\
	This form of the action reminds us of that of a charged particle interacting with magnetic point vertex written in some appropriate variables for facilitating an alternate consistent quantization procedure \cite{JACKIW199083}. Also, the corresponding action was shown there to be scale-invariant. We now move on to study the scale transformation properties of the action (\ref{nwe ac}).

		\section{Broken dilatation symmetry and anomalous Ward-Takahashi identities for $0+1$ dimensional noncommutative fields}

		In this section, we study dilatation symmetry on noncommutative space by using the equivalent commutative description. Further we will be deriving the anomalous Ward-Takahashi (W-T) identities and an explicit evaluation of the dilatation anomaly will be carried out in the path integral approach. The methods developed here capture the quantum effects.

		\subsection{Broken dilatation symmetry and non-conserved dilatation charge  }

		In a series of papers, Jackiw \cite{JACKIW199083,Jackiw:1980mm} has shown the existence of time dilatation symmetry for a charged point particle interacting with magnetic point vortex and magnetic monopole at the classical level. On the other hand, in \cite{kamath1992Dilatation} the dilatation symmetry of the non-relativistic Landau problem has been investigated. However it has been found that there is no dilatation symmetry associated with the non-relativistic Landau problem, thus giving rise to a non-conserved dilatation current at the classical level and further a dilatation anomaly at the quantum level. Note that all these problems has a similar form for the Lagrangian but their scaling properties are different. Motivated by this observation, we investigate whether the generalized Landau problem respect the symmetry and if any dilatation anomaly pop up. Now, by inspection we see that the effective action (\ref{nwe ac}) is not invariant under a global time dilatation:
			\begin{equation}\label{scaletrans}
		t\rightarrow t^{'}= e^{-\gamma}t,
		\end{equation}
		where $\gamma$ is a real parameter. Following Jackiw \cite{JACKIW199083,Jackiw:1972cb}, the form variation of the field $q_{i}(t)$ due to infinitesimal dilatation transformation is given by -     
		\begin{equation}
		\delta_{0} {q}_{i}=\gamma(t\dot{{q}}_{i}-\frac{1}{2}{q}_{i});
		\label{doc}
		\end{equation}
		where $\delta_{0}q_{i}(t)=q^{'}_{i}(t)-q_{i}(t)$ is the functional change. It is now possible to write the non-conserved dilatation generator \cite{Coleman:1985rnk} from a Noether analysis of (\ref{effctive}). To derive the generator of the broken dilatation symmetry, it will be convenient to allow the constant $\gamma$ in (\ref{doc}) to depend on
		time: $\gamma=\gamma(t)$ i.e., infinitesimal local  variation viz. 
		\begin{equation}
		\delta^{L}_{0} {q}_{i}(t)=\gamma(t)\mathcal{Q}q_{i}(t),
		\label{dla}
		\end{equation}
		where $\mathcal{Q}q_{i}(t)=(t\frac{\partial}{\partial t}-\frac{1}{2})q_{i}(t).$ The change in the action resulting from these transformations (\ref{doc}) is
		\begin{align}
		\delta^{L}_{0}{S}_{eff} & = \int dt\bigg[\frac{\partial L_{eff}}{\partial \dot{q}_{i}}\dot{\gamma}\mathcal{Q}q_{i}(t)+\gamma(t)\bigg(\frac{\partial L_{eff}}{\partial \dot{q}_{i}}\mathcal{Q}\dot{q}_{i}(t)+\frac{\partial L_{eff}}{\partial q_{i}}\mathcal{Q}q_{i}(t)\bigg)\bigg]\\
		& =\int dt\bigg[p_{i}\dot{\gamma}\mathcal{Q}q_{i}+\gamma(t)\bigg(\frac{d}{dt}(tL_{eff})+k_{1}{q}_{i}^2+\frac{m_{B}\theta k_{1}}{2\hbar}\epsilon_{ij}q_{j}\dot{q}_{i}\bigg)\bigg],
		\end{align}
		where $p_{i}=\frac{\partial L_{eff}}{\partial \dot{q}_{i}}$ is the canonical momentum. Here, we choose $\gamma(t)$ to decay asymptotically so that one can safely discard the surface
		term 
		and the action $S_{eff}$ then changes as :
		\begin{align}
		\delta^{L}_{0} {S}_{eff} & =\int dt \gamma(t)\bigg( k_{1}{q}_{i}^2+m_{B}k_{1}{\frac{\theta}{2\hbar}}\epsilon_{ij}q_{j}\dot{{q}}_{i}-\frac{d D(t)}{dt}\bigg),
		\label{nd}
		\end{align}
		where $D(t)=tH_{eff}-\frac{1}{2}(q_{i}p_{i})$, and $H_{eff}=\dot{q}_ip_{i}-L_{eff}.$\\
		The expression (\ref{nd}) holds for any field configuration $q_{i}(t)$ with the specific change $\delta ^{L}_{0}q_{i}(t)$. However, when $q_{i}(t)$  obeys the classical equations of motion then $\delta S_{eff}=0$ for any $\delta q_{i}$
		including the symmetry transformation (\ref{dla}) with $\gamma(t)$ a function of time. This
		means that at the on-shell level, we have the  non-conserved dilatation charge ($D$) corresponding to the (broken) dilatation symmetry as -  
			\begin{equation}
		\frac{d D(t)}{dt}= \Delta_{0}(t)
		\label{ncd}
		\end{equation}
		where $\Delta_{0}(t):=\Delta_{0}(q_{i},\dot{q}_{i};t)=k_{1}({q}_{i}^{2}+m_{B}\frac{\theta}{2\hbar}\epsilon_{ij}q_{j}\dot{q}_{i}).$\\
		
		Thus, we identify the non conserved dilatation charge $D$ which is nothing but the generator of the infinitesimal global transformation (\ref{doc}). Here, we observe that scale invariance is broken explicitly by the presence of parameter $k_{1}$ in the action (\ref{nwe ac}); the
		dilatation charge $ D$ acquires a non vanishing time derivative (\ref{ncd}). If $k_{1}$  vanishes, the scale transformation (\ref{scaletrans}) has no effect on the dynamics and therefore corresponds to a symmetry at the classical level.
		
		\subsection{Anomalous Ward-Takahashi Identities}

		Now we will move ahead and discuss about the path integral formulation of Ward-Takahashi
		(W-T) identities associated with the action described in \eqref{nwe ac}. At zeroth order, it represents the quantum mechanical version of Noether's current conservation theorem. Here in this section, we will explicitly derive the W-T identities upto 2nd order and during the course of derivation of these W-T identities, it will be evident that the contribution from Jacobian to be non-trivial. It is precisely this contribution from the Jacobian which gives rise to the anomalous terms in the Ward identities. The anomalous term arising here is eventually regularized using Fujikawa's prescription \cite{Fujikawa:2004cx}.\\

		The starting point is to consider the generating functional for connected Green's functions in $0+1$ dimensional QFT defined in equivalent commutative description (unconstrained variables) $\eqref{connec}$. Here we switch on the source adiabatically and thus the higher 
		order derivatives of $J_{R}^{(1)}$ and $J_{R}^{(2)})$ can be left out (we also assume $J_{R}^{(i)}$ to vanish at boundaries), leaving ({\ref{connec}}) to be :
		\begin{normalsize}
			\begin{equation}
			 \begin{alignedat}{1} & Z\left[J^{(1)}_{R},J^{(2)}_{R}\right] 
& =\int\mathcal{D}X(t) ~~~e^{\frac{i}{\hbar}\left[ S_{eff}[X(t)]+\lambda\int^{\infty}_{-\infty} dt J_{R}^{T}.X+{m_{B}}(\frac{\lambda \theta}{2\hbar})^{2}\int^{\infty}_{-\infty}dt (\frac{1}{2}J_{R}^{T}.J_{R}+\frac{1}{2}\frac{i\theta m_{B}}{\hbar}J^{T}_{R}\sigma_{y}\dot{J}_{R})\right]}
				\end{alignedat}
				\label{cik}
			\end{equation}
		\end{normalsize}
	Note that in this expression, we have rewritten the action  $S_{eff}$ as,
		\begin{equation}
			{S}_{eff}[X(t)]=\int^{\infty}_{-\infty}dt~ X^{T}(t)\mathfrak{R}X(t)
		\end{equation}
		by defining the column vectors,
		\begin{equation}
			\begin{alignedat}{1}X(t) & =\frac{1}{\sqrt{2}}\begin{pmatrix}
					{q}_{1}(t) && {q}_{2}(t)
				\end{pmatrix}^T;\\
				J_{R}(t) & ={\sqrt{2}}\begin{pmatrix}
					{J}^{(1)}_{R}(t) && {J}^{(2)}_{R}(t)
				\end{pmatrix}^{T},
			\end{alignedat}
			\label{ps2}
		\end{equation}
	and the differential operator $\mathfrak{R}$ -
		\begin{align}
			\mathfrak{R} & :=\begin{pmatrix}
				-(m_{B}\frac{d^2}{dt^2}+k_{1})&& m_{B}k_{1}\frac{\theta}{\hbar}\frac{d}{dt}\\
				-m_{B}k_{1}\frac{\theta}{\hbar}\frac{d}{dt}&&-(m_{B}\frac{d^2}{dt^2}+k_{1})
			\end{pmatrix}\\
			& = -(m_{B}\frac{d^2}{dt^2}+k_{1})I+im_{B}\sigma_{y}k_{1}\frac{\theta}{\hbar}\frac{d}{dt}
		\end{align}
		Here $\mathfrak{R} $ is itself hermitian and its eigenvalues are being considered as discrete here  -
		\begin{equation}
			\mathfrak{R} \phi_{k}(t)=\lambda_{k}\phi_{k}(t),
		\end{equation}
		where the eigenfunctions $\phi_{k}(t)$ satisfy the usual orthogonality and completeness conditions:
		\begin{equation}
			\begin{alignedat}{1}\int dt \phi_{k}^{\dagger}(t)\phi_{j}(t) & =\delta_{kj};\\
				\sum_{k=1}^{\infty}\phi_{k}(t)\phi^{\dagger}_{k}(t^{'}) & =\delta(t-t^{'})\mathbb{I}_{2}.
			\end{alignedat}
			\label{ps1}
		\end{equation}
		In order to properly specify the functional measure, the ($0+1$) dimensional field $X(t)$ are expanded in terms of these eigenfunctions, 
		\begin{equation}
			X(t)=\sum_{k=1}^{\infty} a_{k}\phi_{k}(t)
			\label{fey}
		\end{equation}
		Now, the functional-integral measure is then defined as,
		\begin{equation}
			\mathcal{D}X=\prod_{k=1}^{k=\infty} da_{k}=\mathcal{D}a
		\end{equation}
		To obtain the Ward identities, we study the behaviour of the commutative equivalent action (\ref{nwe ac})
		and the measure defined in (\ref{ci}) under infinitesimal transformation (\ref{dla}) of the field $q_{i}$ specified by
		a local parameter $\gamma(t).$ Therefore, under the local transformation -
		\begin{equation}
			X(t)\rightarrow X'(t)=X(t)+\delta^{L}_{0} X(t),
			\label{lie}
		\end{equation}
		with $\delta^{L}_{0} X(t)=\gamma(t)(t\frac{\partial}{\partial t}-\frac{1}{2})X(t),$ the expansion coefficients of (\ref{fey}) change to-
	\begin{align}
			a_{j}\rightarrow	a_{j}' =a_{j}+\sum_{k}a_{k}\int dt\gamma(t)\phi^{\dagger}_{j}(t)(t\frac{\partial}{\partial t}-\frac{1}{2})\phi_{k}(t)).
			\label{jaco}
		\end{align}
		Thus, the measure changes as \cite{PhysRevLett.44.1733, PhysRevD.21.2848} -
		\begin{equation}
			\mathcal{D}X\rightarrow \mathcal{D}X^{'}=\prod_{j=1}^{j=\infty} da^{'}_{j}=(\det c_{jk})\prod_{k=1}^{k=\infty} da_{k}=e^{Tr (ln \; c_{jk})}\mathcal{D}X
			\label{br}
		\end{equation}
	    where $det( c_{jk})$ can be read as the Jacobian of the transformation (\ref{jaco}), whereas, the transformation matrix $c_{jk}$ can be obtained by,
		\begin{equation}
			c_{jk}=\frac{\partial a'_{j}}{\partial a_{k}}= \delta_{jk}+\int dt \gamma(t)\phi_{j}^{\dagger}(t)(t\frac{\partial}{\partial t}-\frac{1}{2})\phi_{k}(t)
		\end{equation}
		Now, the above generating functional of connected Green's functions (\ref{cik}) may be rewritten as,
	\begin{normalsize}
			\begin{equation}
				\begin{alignedat}{1}
				& Z\left[J^{(1)}_{R},J^{(2)}_{R}\right]
					=\int\mathcal{D}X'(t) ~~~e^{\frac{i}{\hbar}\int^{\infty}_{-\infty} dt\left[ X^{'T}\mathfrak{R}X^{'}+\lambda J_{R}^{T}.X^{'}+{m_{B}}(\frac{\lambda \theta}{2\hbar})^{2}\frac{1}{2} (J_{R}^{T}.J_{R}+\frac{i\theta m_{B}}{\hbar}J^{T}_{R}\sigma_{y}\dot{J}_{R})\right]}\\
					& 
					=\int e^{Tr(ln \; c_{jk})}\mathcal{D}X ~~~e^{\frac{i}{\hbar}\int^{\infty}_{-\infty} dt\left[ X^{T}\mathfrak{R}X+\lambda J_{R}^{T}.X+{m_{B}}(\frac{\lambda \theta}{2\hbar})^{2}\frac{1}{2} (J_{R}^{T}.J_{R}+\frac{i\theta m_{B}}{\hbar}J^{T}_{R}\sigma_{y}\dot{J}_{R})\right]}\\
					&~~~~~~~~~~~~~~~~~~~~~~~~~~~~~~~~~~~~~~~~~~~~~~~~~~\times ~e^{\frac{i}{\hbar} [\delta^{L}_{0}S_{eff}+\lambda\int_{-\infty}^{\infty}dt J^{T}_{R}(t)\delta^{L}_{0} X(t)]},
				\end{alignedat}
				\label{ci}
			\end{equation}
		\end{normalsize}
		The first equality here is a triviality: we have simply re-labelled $ X(t)$  by $X^{'}(t)$ as a dummy variable in
		the functional integral (\ref{cik}). The second equality is nontrivial and uses the assumed transformations (\ref{lie}) and (\ref{br}). For infinitesimal $\gamma(t)$ the generating functional transforms as,
		\begin{align}
			& Z\left[J^{(1)}_{R},J^{(2)}_{R}\right] 
			= \int\mathcal{D}X(t) ~e^{\frac{i}{\hbar}\int^{\infty}_{-\infty} dt\left[ X^{T}\mathfrak{R}X+\lambda J_{R}^{T}.X+{m_{B}}(\frac{\lambda \theta}{2\hbar})^{2}\frac{1}{2} (J_{R}^{T}.J_{R}+\frac{i\theta m_{B}}{\hbar}\epsilon_{ij}J^{T}_{R}\sigma_{y}\dot{J}_{R})\right]}\nonumber\\
			&~~~~~\times e^{\frac{i}{\hbar}[\delta^{L}_{0}S_{eff}+\int_{-\infty}^{\infty}dt (\lambda J^{T}_{R}\delta_{0}^{L} X-i\hbar\gamma(t)A(t))]}
			\label{dbc}
		\end{align}
		where,
		\begin{equation}
			A(t)=\sum_{k}\phi_{k}^{\dagger}(t)(t\frac{\partial}{\partial t}-\frac{1}{2})\phi_{k}(t).
			\label{Anf}
		\end{equation}
		This $A(t)$ is referred to as the quantum  anomaly term. Now reverting back to our previous co-ordinates and performing the Taylor series expansion of the exponential factor in (\ref{dbc}) upto terms of $\mathcal{O}(\gamma)$, we obtain :-
		\begin{align}\label{R}
			& Z\left[J^{(1)}_{R},J^{(2)}_{R}\right]=\int \mathcal{D}q_{1}(t)\mathcal{D}q_{2}(t)e^{\frac{i}{\hbar}\left[ S_{eff}+\int^{\infty}_{-\infty} dt [\lambda J_{R}^{(i)}q_{i}+{m_{B}}(\frac{\lambda \theta}{2\hbar})^{2} ({J_{R}^{(i)}}^2+\frac{\theta m_{B}}{\hbar}\epsilon_{ij}J_{R}^{(i)}\dot{J_{R}}^{(j)})\right]}\nonumber\\
			&~~~~~~~~~~~~~~  \times \left(1+\frac{i}{\hbar} \int dt \gamma(t)(\Delta_{0}-\frac{d D}{dt}
			+\lambda J_{R}^{(i)}(t)\mathcal{Q}{q}_{i}(t) -i\hbar A(t))+\mathcal{O}(\gamma^{2})\right), 
		\end{align}
		where in the last line we have made use of eq.(\ref{nd}). The W-T identities are summarized by the variational derivative
		(the change of integration variables does not change the integral
		itself). In particular the variation with respect to $\gamma(t)$ must vanish, since it holds for any value of $\gamma$, i.e.
		\begin{equation}
			\frac{\delta Z[J^{(1)}_{R},J^{(2)}_{R}]}{\delta \gamma(t')}=0
		\end{equation}
		Thus Taylor expanding the exponential in (\ref{R}) containing source terms and then making their variation w.r.t. $\gamma(t')$ go to zero, we have (where $|t'|<\infty$):- 			
		\begin{align}
			& 0 =  \int \mathcal{D}q_{1}(t)\mathcal{D}q_{2}(t)e^{\frac{i}{\hbar}S_{eff}}\nonumber \\
			& \times (1+\frac{i \lambda}{\hbar}\int dt_{1} J_{R}^{(i)}(t_{1})q_{i}(t_{1})-\frac{{\lambda}^2}{2{\hbar}^2}\int\int dt_{1}dt_{2}J^{(i)}_{R}(t_{1})J^{(k)}_{R}(t_{2})q_{i}(t_{1})q_{k}(t_{2})\nonumber\\
			& +\frac{{\lambda}^2}{2{\hbar}^2}\int\int dt_{1}dt_{2}\frac{im_{B}}{2\hbar}\delta(t_{1}-t_{2})\left[(J_{R}^{(i)}(t_{1}))^2+\frac{\theta m_{B}}{\hbar}\epsilon_{ij}J_{R}^{(i)}(t_{1})\dot{J_{R}}^{(j)}(t_{2})\right]+........)\nonumber  \\
			& \times \frac{i}{\hbar}\left(\Delta_{0}-\frac{d D}{dt^{'}}
			+ {\lambda}J_{R}^{(i)}(t^{'})\mathcal{Q}{q}_{i}(t^{'}) -i\hbar A(t^{'})\right),
		\end{align}\\\\
		Rearrangement of the previous expression yields:			
		\begin{align}\label{m}
			& 0 =  \int \mathcal{D}q_{1}(t)\mathcal{D}q_{2}(t)e^{\frac{i}{\hbar}S_{eff}}(\frac{i}{\hbar}  (\Delta_{0}(t')-\frac{d D}{dt'}-i\hbar A(t'))\nonumber\\& +\frac{i\lambda}{\hbar}[\frac{i}{\hbar}\int dt_{1}J^{(i)}_R(t_{1})q_{i}(t_{1})(\Delta_{0}(t')-\frac{d D}{dt'}-i\hbar A(t'))\nonumber\\
			& +\int dt_{1}J^{(i)}_{R}(t_{1})\mathcal{Q}q_{i}(t_{1})\delta(t_{1}-t')]\nonumber\\
			& -\frac{{\lambda}^2}{2{\hbar}^2}[\int\int dt_{1}dt_{2}J^{(i)}_{R}(t_{1})J^{(k)}_{R}(t_{2})q_{i}(t_{1})\mathcal{Q}q_{k}(t_{2})\delta(t_{2}-t')\nonumber\\
			&+\int\int dt_{1}dt_{2}J^{(i)}_{R}(t_{1})J^{(k)}_{R}(t_{2})\mathcal{Q}q_{i}(t_{1})q_{k}(t_{2})\delta(t_{1}-t')\nonumber\\
			& +\frac{i}{\hbar}\int\int dt_{1}dt_{2}J^{(i)}_{R}(t_{1})J^{(k)}_{R}(t_{2})q_{i}(t_{1})q_{k}(t_{2})({\Delta}_{0}(t')-\frac{d D}{dt'}-i\hbar A(t'))\nonumber\\
			&-\int\int dt_{1}dt_{2}\frac{im_{B}}{2{\hbar}}\delta(t_{1}-t_{2})((J^{(i)}_{R}(t_{1}))^2+\frac{\theta m_{B}}{\hbar}\epsilon_{ij}J^{(i)}_{R}(t_{1})\dot{J}^{(j)}_R(t_{2}))\nonumber\\
			&\times \frac{i}{\hbar}({\Delta}_{0}(t')-\frac{d D}{dt'}-i\hbar A(t'))]+...............)
		\end{align}
		The different orders of Ward-Takahashi identities are obtained in terms of mean values of dynamical quantities \footnote{where the mean values are defined as: $\left\langle T^*[............]\right\rangle=\int\mathcal{D}q_{1}(t)\mathcal{D}q_{2}(t)[............]e^{\frac{i}{\hbar}S_eff}$ apart from the aforementioned normalization factor.} by taking the variation of right hand side of (\ref{m}) with respect to the source term i.e. $\frac{\delta}{\delta J^{(i)}_{l}(t'_{1})}\frac{\delta}{\delta J^{(m)}_{R}(t'_{2})}.....$ and then setting $J^{(l)}_{R}(t)$'s to $0$.
		\\
		\\
		Thus the zeroth-order W-T identity:-
		\begin{normalsize}
			\begin{align}\label{af}
				\frac{d}{dt'}\left\langle D(t')\right\rangle=\left\langle\Delta_{0}(t')\right\rangle-i\hbar\left\langle A(t')\right\rangle
			\end{align} 
		\end{normalsize}
		First-order W-T identity:-
		\begin{normalsize}
			\begin{align}\label{aff}
				\frac{d}{dt'}\left\langle T^{*}[D(t')q_{i}(t'_{1})]\right\rangle  = & \left\langle T^{*}[\Delta_{0}(t')q_{i}(t'_{1})]\right\rangle-i\hbar \left\langle T^{*}[A(t')q_{i}(t'_{1})]\right\rangle\nonumber \\
				&-i\hbar\delta(t'_{1}-t')\left\langle\mathcal{Q}{q}_{i}(t_{1})\right\rangle
			\end{align}
		\end{normalsize}
		Second-order W-T identity:-
		\begin{normalsize}
			\begin{align}\label{afff}
				\frac{d}{dt'}\left\langle T^{*}[D(t'){q}_{k}(t'_{2}){q}_{i}(t'_{1}))]\right\rangle & =\left\langle T^{*}[\Delta_{0}(t'){q}_{k}(t'_{2}){q}_{i}(t'_{1})]\right\rangle\nonumber\\
				& -i\hbar\delta(t'_{1}-t')\left\langle T^{*}[{q}_{k}(t'_{2})\mathcal{Q}{q}_{i}(t'_{1})\right\rangle\nonumber\\
				& -i\hbar\delta(t'_{2}-t')\left\langle T^{*}[\mathcal{Q}{q}_{k}(t'_{2}){q}_{i}(t'_{1})]\right\rangle\nonumber\\
				& -i\hbar \left\langle T^{*}[ A(t'){q}_{k}(t'_{2}){q}_{i}(t'_{1})]\right\rangle 
			\end{align}
		\end{normalsize}
		Similarly one can derive the higher-orders Ward-Takahashi identities. Particularly, the zeroth-order W-T identity tells us that the dilatation symmetry is explicitly broken due to presence of the spring-constant $k_{1}$ in the Lagrangian (\ref{effctive}), which is consistent with that from Noether's prescription (\ref{ncd}). But at the quantum level, there is also an additional second term on the right which is the contribution from the existence of anomaly at the quantum level; it is the anomaly term $A(t)$ which is responsible for modifying the rate of change of dilatation charge $D(t).$ The higher-order identities also indicate this anomalous behavior up to "contact terms".\\

			It is worthwhile to note that in this derivation of W-T identities, the  ill-defined expression of the anomaly term in (\ref{Anf}) apparently seems divergent. This is because, at each time $t$ we are summing over an infinite number of modes $\phi_{k}(t)$, however it is possible to give a meaning to this expression by the method of regularization and subsequently extract out a non divergent result as we will see in the next section.

		\section{Computation of the dilatation anomaly: Fujikawa's method}
		
		Now we are in a position to compute the anomalous term in W-T identities. Following Fujikawa's prescription \cite{PhysRevD.21.2848}, the anomaly can be regularized by correcting each contribution from $\phi_{k}(t)$ by a factor $e^{-\lambda_{k}^2/M^2}$ as below -
		
		\begin{equation}
		A(t)\rightarrow A_{Reg.}(t):=\lim_{M\to \infty}\sum_{k} \phi_{k}^{\dagger}(t)(t\frac{\partial}{\partial t} -\frac{1}{2})\phi_{k}(t) e^{-\lambda_{k}^2/M^2}
		\label{reg}
		\end{equation}
		Here $\phi_{k}(t)\phi^{\dagger}_{k}(t)$ is an ill-defined object and can be interpreted consistently by first separating time points  and then taking the coincident limit at  a suitable step. A simple manipulation therefore yields:

		\begin{align}\label{reganomaly1}
		A_{Reg.}(t)& =  \lim_{M\to \infty}\lim_{t\to t'}~ tr \sum_{k} (t\frac{\partial}{\partial t} -\frac{1}{2})e^{-\mathfrak{R}^2/M^2}\phi_{k}(t)\phi_{k}^{\dagger}(t')\nonumber \\
		& =  \lim_{M\to \infty}\lim_{t\to t'}~ tr
		 \bigg[(t\frac{\partial}{\partial t} -\frac{1}{2})e^{-\mathfrak{R}^2/M^2} \delta(t-t')\mathbb{I}_{2}\bigg]\nonumber \\
		& = \lim_{M\to \infty}\lim_{t\to t'}~ tr \int_{-\infty}^{\infty} \frac{dz}{2\pi}(t\frac{\partial}{\partial t} -\frac{1}{2})e^{-\mathfrak{R}^2/M^2}e^{iz(t-t')}\mathbb{I}_{2}
		\end{align}
		 Note that in above, tr refers to only the $2\times 2$ matrix indices whereas the Tr appearing earlier in (\ref{br}) refers to both functional and matrix indices. Here we have used the completeness relation of $\phi_{k}(t)$ (\ref{ps1}) and in the last step, the integral representation of $\delta$-function has been used. Now we define ${\theta}'=\frac{\theta}{2\hbar}$ to be used from now onwards in order to remove any $\hbar$ dependency of $\theta$ rendering $\theta'$ a classical parameter. The anomaly expression (\ref{reganomaly1}) after evaluating $\lim_{t\to t'}e^{-\frac{\mathfrak{R}^2}{M^2}}e^{iz(t-t')}$ and then taking the limit $t\to t'$ becomes -

		\begin{equation}
		A_{Reg.}(t)=\lim_{M\to\infty}~tr~\int_{-\infty}^{\infty}\frac{dz}{2\pi}(izt-\frac{1}{2})e^{-a\bf{I}+b\bf{\sigma_{y}}}\mathbb{I}_{2},
		\end{equation}
		where
		\begin{align*}
		a & =\frac{1}{M^2}(m_{B}^2{z^4}+4m_{B}^2k_{1}^2{\theta}'^2{z^2}-2m_{B}k_{1}z^2+k_{1}^2)\\
		b & = \frac{4m_{B}k_{1}{\theta}'}{M^2}(m_{B}z^3-k_{1}z)
		\end{align*}
		Using $e^{-a\bf{I}}=e^{-a}\bf{I}$, $e^{b\bf{\sigma_{y}}}=\bf{I}\cosh{b}+\bf{\sigma_{y}}\sinh{b}$ and completing the trace operation, we have :-
		\begin{equation}
		A_{Reg.}(t)=2\lim_{M\to\infty}\int_{-\infty}^{\infty}\frac{dz}{2\pi}(izt-\frac{1}{2})e^{-a}\cosh{b}
		\label{frg}
		\end{equation}
		Since $e^{-a}\cosh{b}$ is an even function in $z$, the first term in the integral doesn't contribute. Therefore, the regularized version of the anomaly term (\ref{frg}) is rewritable in a much simpler form as,
		 
		\begin{equation}
		A_{Reg.}(t)=-\lim_{M\to\infty}\frac{1}{2\pi}(I_{1}+I_{2}),
		\end{equation}
		where the integrals $I_1$ and $I_2$ are given by -
		\begin{align*}
		I_{1}=\int_{0}^{\infty}dze^{-a+b}\\
		I_{2}=\int_{0}^{\infty} dz e^{-a-b}
		\end{align*}
		For simplicity in notation, let $z^2-2k_{1}{\theta}'z=s$,~ $\frac{m_{B}^2}{M^2}=p$,~ $p{\omega'}^2=q$ (where ${\omega'}^2=k_{1}^2{\theta}'^2+\frac{k_{1}}{m_{B}}$) and $r=\sqrt{s+k_{1}^2{\theta}'^2}$. The integral $I_1$ becomes :
		\begin{align}
		I_{1} & =\int_{0}^{\infty}\frac{ds}{2\sqrt{s+k_{1}^2{\theta}'^2}} \exp{-p(s-{\omega}^2)^2}\nonumber\\
		& = \int_{k_{1}^2{\theta}'^2}^{\infty}\frac{dr}{2\sqrt{r}}\exp{-p(r-\omega'^2)^2}\nonumber\\
		& = e^{-p{\omega'}^4}\int_{k_{1}^2{\theta}'^2}^{\infty}\frac{dr}{2\sqrt{r}}e^{-pr^2+2qr}
		\end{align}
		Following similar steps, one can show :-
		\begin{equation}
		I_{2} = e^{-p{\omega'}^4}\int_{k_{1}^2{\theta}'^2}^{\infty}\frac{dr}{2\sqrt{r}}e^{-pr^2+2qr}
		\end{equation}
		Thus the regularised anomaly factor $A_{Reg.}(t)$ simplifies to -
		\begin{align}
		A_{Reg.}(t) & =-\lim_{M\to\infty}\frac{1}{2\pi}e^{-p{\omega'}^4}\int_{k_{1}^2{\theta}'^2}^{\infty}dr \;{r}^{-1/2}e^{-pr^2+2qr}\nonumber\\
		& =- \frac{1}{2\pi}\lim_{M\to\infty}e^{-p{\omega'}^4}\left[\int_{0}^{\infty}dr \;{r}^{-1/2}e^{-pr^2+2qr}-\int_{0}^{k_{1}^2{\theta}'^2}dr \;{r}^{-1/2}e^{-pr^2+2qr}\right] \label{finalregularised}
		\end{align}
		Upon taking the limit $M\to\infty$, the 2nd integral in (\ref{finalregularised}) is convergent and gives a finite result of $2k_{1}\theta '$. The first integral can be evaluated as in \cite{kawa:2004cx}:
		\begin{equation}
		\lim_{M\to\infty}e^{-p{\omega'}^4}\int_{0}^{\infty}\frac{dr}{\sqrt{r}}e^{-pr^2+2qr}=\lim_{M\to\infty}(2p)^{-\frac{1}{4}}\Gamma\left(\frac{1}{2}\right)D_{-\frac{1}{2}}(y)e^{\frac{y^2}{2}}e^{-p{\omega '}^4}
		\end{equation}
		where $y=\frac{p{\omega'}^2}{\sqrt{2p}}$. We can also write $D_{-1/2}(y)$ in terms of $K_{1/4}(y)$, where $D_{-1/2}(y)$ and $K_{1/4}(y)$ represent the parabolic cylindrical functions and modified Bessel function respectively.
		\begin{equation}
		\lim_{M\to\infty}(2p)^{-\frac{1}{4}}\Gamma\left(\frac{1}{2}\right)D_{-\frac{1}{2}}(y)e^{\frac{y^2}{2}}=\lim_{M\to \infty}(2p)^{-\frac{1}{4}}e^{\frac{y^2}{2}}(\frac{1}{4}y^2)^{\frac{1}{4}}K_{1/4}(\frac{1}{4}y^2)
		\end{equation}
		Using $K_{1/4}(x)\approx\frac{1}{2}\Gamma(\frac{1}{4})(\frac{x}{2})^{-\frac{1}{4}}$ as $x\rightarrow 0$, we have :-
		\begin{equation}
		\lim_{M\to\infty}e^{-p{\omega'}^4}\int_{0}^{\infty}\frac{dr}{\sqrt{r}}e^{-pr^2+2qr}=\frac{1}{2}\lim_{M\to\infty} \Gamma\left(\frac{1}{4}\right)p^{-\frac{1}{4}}= \frac{1}{2}\lim_{M\to\infty} \Gamma\left(\frac{1}{4}\right)(M/m_{B})^{1/2}
		\end{equation}
		\begin{equation}
		A_{Reg.}(t)=-\frac{1}{2\pi}\left[ \frac{1}{2}\lim_{M\to\infty} \Gamma\left(\frac{1}{4}\right)(M/m_{B})^{1/2}-2k_{1}{\theta '}\right]
		\label{final}
		\end{equation}

		The first term in the expression is independent of the coupling $k_{1}$ and is the same as in the non-interacting case of our equivalent effective commutative theory, which is usually considered to be non-anomalous and only the second term is independent of the cutoff $M$. Therefore, following conventional wisdom \cite{PhysRevD.39.3672}, we now concentrate our attention to the nontrivial finite contribution coming from (\ref{final}). To obtain a non-divergent anomaly contribution, we need to renormalize the relation (\ref{final}) in free particle theory limit $k_{1}\rightarrow 0$ by simply adding a term $i\hbar \gamma(t)A_{f}$ in the Lagrangian $L_{eff}$ (\ref{dbc}), thus cancelling the divergence in (\ref{final}). Here, $A_{f}$ is equal to the Fujikawa factor for the free particle theory,
		\begin{equation}
		A_{f}=-\frac{1}{4\pi}\lim_{M\to\infty} \Gamma\left(\frac{1}{4}\right)(M/m_{B})^{1/2} 
		\end{equation}
		  Therefore, after renormalization the correct dilatation anomaly is given by :
		\begin{equation}
		A_{renormalized}= A_{Reg.}-A_{f}=\frac{1}{\pi}k_{1}{\theta '}
		\end{equation}        
		Taking into account the above renormalization scheme, only the finite part of the anomaly term contributes to obtain renormalized version of the set of bare Ward identities given in (\ref{af}-\ref{afff}).
One can also notice the fact that the anomalous correction to the W-T identities is a first order in $\theta$, where $\theta=\frac{\hbar c}{eB}$, which suggests that for commutative limit i.e. $\theta\rightarrow 0$, one must have $\hbar\rightarrow 0$, thus implying that the commutative limit coincides with the classical limit. This indeed attests to the fact that in this setting, the correct quantum corrections are fully taken into account only when we switch to the noncommutative scenario.

		\section{Conclusion}
		Firstly, we summarize our key results. By considering a pair of non-relativistic interacting opposite charged particles in a
		region of high constant magnetic field and low mass limit, we show how the Lagrangian of the system can be directly mapped to a harmonic oscillator in noncommutative
		space, referred here as exotic oscillators. A quantum picture is also provided for the above using Hilbert-Schimdt operators, thereby unleashing a physical setting of this abstract formulation of noncommutative quantum mechanics. Therefore, the noncommutativity is given a physical model here unlike in many other works where it is postulated as a fundamental parameter. We then construct an effective path integral action for the noncommutative system
		using coherent states in a method similar to the one followed by Gangopadhyay and
		Scholtz in \cite{PhysRevLett.102.241602}, however we have included the source term here. The effective commutative action derived through path-integral describes a charged particle moving in a constant magnetic field and under the influence of a harmonic oscillator potential - the generalized Landau problem. Also the form of this action is reminiscent of that of a charged particle interacting with point magnetic vortex studied in \cite{JACKIW199083} where scale symmetry exists in contrast to the present case, suggesting that thorough treatment is required for each individual case. 
		We compute the anomalous W-T identities which necessarily includes
		contribution from the Jacobian. This has been unexplored in the literature from the point of view of noncommutative quantum mechanics under scaling transformations. The anomalous term has been calculated and regularised following
		Fujikawa's method.
		 
		We now make certain pertinent observations in connection with the problem. It is to be pointed out that our investigation suggests the emergence of scale anomalies in interacting quantum Hall systems as a natural consequence of impurity interactions \cite{doi:10.1142/S0217751X04018099} upon quantization in large magnetic field limit. This is because we have built our noncommutative model starting from the Lagrangian (\ref{hall}). We also mention that the computation of anomalous correction in W-T identities is independent of the status of noncommutativity parameter $\theta$ - whether it being an independent fundamental constant or a derived one as is the case here. The anomalous correction is directly proportional to the
		noncommutative parameter $\theta$, which suggests that quantum corrections arise due to 
		the noncommutativity of spatial coordinates. One can show, by considering the noncommutativity parameter to be fundamental, that the
		dilatation current for a harmonic oscillator in commutative space is not inflicted with
		an anomaly upon quantization. Also in this regard, we mention that the anomaly contribution in the non-conservation of dilatation current (i.e. in zeroth-order W-T identity) (\ref{af}) being imaginary can be traced to the fact that domain of definition of the Hamiltonian does not remain invariant under quantum corrections \cite{PhysRevD.66.125013}. It is also worth mentioning that the effect of noncommutativity can be traced back to a curved momentum space \cite{Welling_1997} and our analysis hints at an emergence of scale anomaly in such a curved momentum space. In the end, we want to point out a conceptual resemblance of the present problem with the massive scalar $\phi^4$ theory where a scale anomaly appears upon quantization besides the explicit broken charge owing to the mass of field in the sense that here also there is explicit breaking of scale symmetry due to the spring constant with an additional anomalous correction occurring upon quantization. This feature motivates an ambitious task of probing the effects of Planck scale in "contact" interacting two-dimensional ultracold Bose gas by starting from a noncommutative (2+1) massive relativistic scalar $\phi^4$ theory and then taking some suitable non-relativistic limit \cite{PhysRevLett.105.095302, PhysRevD.46.5474}.

	Finally, our present work can be extended in the following directions: (i) Very recently, in \cite{PhysRevA.100.023601}, it has been shown that there exists a connection between conformal symmetry breaking and production of entropy and also in a slightly different perspective in \cite{Cavalcante:2014rla}. We intend to investigate along this line as deformed quantum system has been shown to produce entropy in \cite{Pal:2020jv}. (ii) Another important avenue of further study will be to look into the modified Virial theorem in the Euclidean version theory of the scale anomaly studied here, so that one can find the pressure-energy relation in this system. It is generally believed that anomalies are UV effects whereas Berry phase is an infrared effect - here it seems interesting to explore any mixing between anomalies and geometrical phases in noncommutative spaces in the adiabatic approximation as hinted from the following studies \cite{PhysRevD.97.016018, Minwalla_2000, PhysRevA.102.022231}.
		 
		 \section*{Acknowledgements}
		 
		 We would like to extend our gratitudeness to Prof. Biswajit Chakraborty for a careful reading of the manuscript and for his questions and critical comments on the paper. One of the authors (P.N.) thanks Professor Kazuo Fujikawa for a correspondence and for enlightening him about quite a few subtle points in connection with this paper. Also he has benefited from conversations with Frederik G. Scholtz, Alexei Deriglazov. The authors thank V.P. Nair for various insightful comments on an earlier draft which helped us to improve the manuscript. We sincerely thank Prof. Rabin Banerejee and Debasish Chatterjee for introducing us to the subject. Special thanks are also due to Prof. A. P. Balachandran for his constant encouragement. S. S. is funded by Jagadis Bose National Science Talent Search (JBNSTS) scholarship and S. K. P. thanks UGC-India for providing financial assistance in the form of fellowship during the course of this work.

						\bibliographystyle{ieeetr}
						\bibliography{torr}
						
					\end{document}